\documentclass{IEEEtran}
\usepackage{cite}
\usepackage{amsmath,amssymb,amsfonts}
\usepackage{algorithmic}
\usepackage{graphicx}
\usepackage{textcomp}
\usepackage{subfig}
\usepackage{dblfloatfix}
\usepackage[table]{xcolor}
\usepackage{tkz-graph}

\newtheorem{remark}{Remark}
\newtheorem{definition}{Definition}

\GraphInit[vstyle = Shade]
\tikzset{
  LabelStyle/.style = { rectangle, rounded corners, draw,
                        minimum width = 2em, fill = yellow!50,
                        text = red, font = \bfseries },
  VertexStyle/.append style = { inner sep=5pt,
                                font = \normalsize\bfseries},
  EdgeStyle/.append style = {->, bend left} }
\usetikzlibrary {positioning}

\newcommand*{\new}{\textcolor{black}}

\def\BibTeX{{\rm B\kern-.05em{\sc i\kern-.025em b}\kern-.08em
    T\kern-.1667em\lower.7ex\hbox{E}\kern-.125emX}}
\begin{document}
\title{\new{Learning-based hierarchical control of water reservoir systems}}
\author{Pauline Kergus$^{1}$, Simone Formentin$^{2}$, Matteo Giuliani$^{2}$, and Andrea Castelletti$^{2}$
\thanks{This project was partially supported by the Italian Ministry of University and Research under the
PRIN’17 project "Data-driven learning of constrained control systems", contract no. 2017J89ARP\new{, and by Regione Lombardia under the project ADDApt.}}
\thanks{$^{1}$ Department of Automatic Control, Lund University, Sweden (e-mail: pauline.kergus@control.lth.se) }
\thanks{$^{2}$ Dipartimento di Elettronica, Informazione e Biongegneria, Politecnico di Milano, Italy (e-mail: simone.formentin@polimi.it, matteo.giuliani@polimi.it, andrea.castelletti@polimi.it)}
}

\bibliographystyle{IEEEtran}

\maketitle

\begin{abstract}
The optimal control of a water reservoir systems represents a challenging problem, due to uncertain hydrologic \new{inputs} and the need to adapt to changing environment and varying control objectives. In this work, we propose a real-time learning-based control strategy based on a hierarchical predictive control architecture. Two control loops are \new{implemented}: the inner loop is aimed to make the overall dynamics similar to an assigned linear \new{through data-driven control design}, then the outer economic model-predictive controller compensates for model mismatches, enforces suitable constraints, and boosts the tracking performance. The effectiveness of the proposed approach as compared to traditional dynamic programming strategies is illustrated on an accurate simulator of the Hoa Binh reservoir in Vietnam. \new{Results show that the proposed approach performs better than the one based on stochastic dynamic programming.}
\end{abstract}


\section{Introduction}
\label{sec:introduction}
On a global scale, population and economic growth result in an increasing energy demand. At the same time, climate change and growing populations pressure freshwater resources \cite{mcdonald2011urban}. Hydropower dams could constitute a response to these big energy challenges since they provide a wide range of benefits such as reduced fossil fuel consumption, irrigation and urban water supply. \new{Yet, building new dams implies} substantial financial and environmental costs and, as explained in \cite{ansar2014should}, many large storage projects worldwide are failing to produce the level of benefits that would economically justify their development. Therefore, operating existing infrastructures more efficiently, rather than planning new ones, \new{represents an opportunity for these facilities to fulfill their potential, which is a critical challenge}. New operating policies should be able to adapt release decisions to uncertain hydrologic conditions and to \new{evolving} objectives such as growing water demands \cite{soncini2007integrated}.

This problem has received much attention from different research fields since the 70s but, as explained in \cite{castelletti2008water}, it remains challenging for different reasons: water reservoirs models are highly non-linear and multiple and conflicting interests are at stakes, usually formulated as non-linear and strongly asymmetric objective functions. In addition, the system is affected by strong uncertainties, such as the inflow of water, which cannot be neglected. In the literature, dynamic programming (DP) and its stochastic extension (SDP) are among the most widely used methods for designing optimal operating policies for water reservoirs, see \cite{castelletti2008water}. In practice, the use of SDP is limited by \new{the curse of dimensionality \cite{Bellman1957} and the curse of modelling \cite{Powell2007}.}

In this paper, we propose to control reservoirs operations using a mild knowledge of the system description and a hierarchical learning-based approach using operation data as illustrated in \cite{piga2017direct}. First, a parametric controller is designed to match some desired closed-loop behavior using the Virtual Reference Feedback Tuning approach of \cite{campi2002virtual}. An outer Model Predictive Controller (MPC) is then used as a reference governor, enabling to enforce constraints on the water release and to compensate for possible tracking deficiencies. It follows that the problem of selecting an achievable reference model becomes less critical than in \cite{campi2002virtual} since low-performance models can be employed, and as such they are easily achieved with a simple control structure like PID (Proportional Integral Derivative)-like blocks. 

The above approach shows a number of advantages over traditional SDP: (i) it can be easily implemented online, (ii) being ``model-free" (it does not require a full mathematical description of the \new{controlled} reservoir), it can be easily adapted in real-time to any change in the system or operating conditions, (iii) it can handle time-varying constraints within a receding-horizon rationale. Nonetheless, the method presents a number of tuning knobs, whose selection will be discussed throughout the paper.

The above learning-based approach has already shown its potential in implicit force control for robotics \cite{polverini2019mixed} and position control in mechatronics \cite{piga2019performance}. However, this is the first time such a rationale is applied to a hydropower system, where the overall control objective is formulated in an economic MPC fashion (see \cite{faulwasser2018economic}).

\new{This work is illustrated on the Hoa Binh water reservoir system in Vietnam, for which a simulation model is available along with historical data. The objective is to control the operations of this water reservoir to maximize hydroelectricty production and minimize flooding while taking into account the constraints of the system and the hydrologic inputs. The whole paper demonstrates the effectiveness of the proposed approach on this case study, and compares its performance with classical SDP as used in \cite{castelletti2008water}.}

The remainder of the paper is organized as follows. First, the Hoa Binh case study is presented, followed by a brief summary of the classical \new{SDP} approach typically employed for policy search. The proposed hierarchical approach is then described, starting from the data-driven design of the inner-loop based on the VRFT method \cite{campi2002virtual}, and then introducing the outer economic MPC loop, which plays the role of a reference governor. Simulation results are then reported and compared with SDP. Concluding remarks, along with outlooks for future research, are presented in the last section.

\section{Preliminaries}
\label{sec:preliminaries}
\subsection{The Hoa Binh case study}

\begin{figure}[h]
    \centering
    \includegraphics[width=0.45\textwidth]{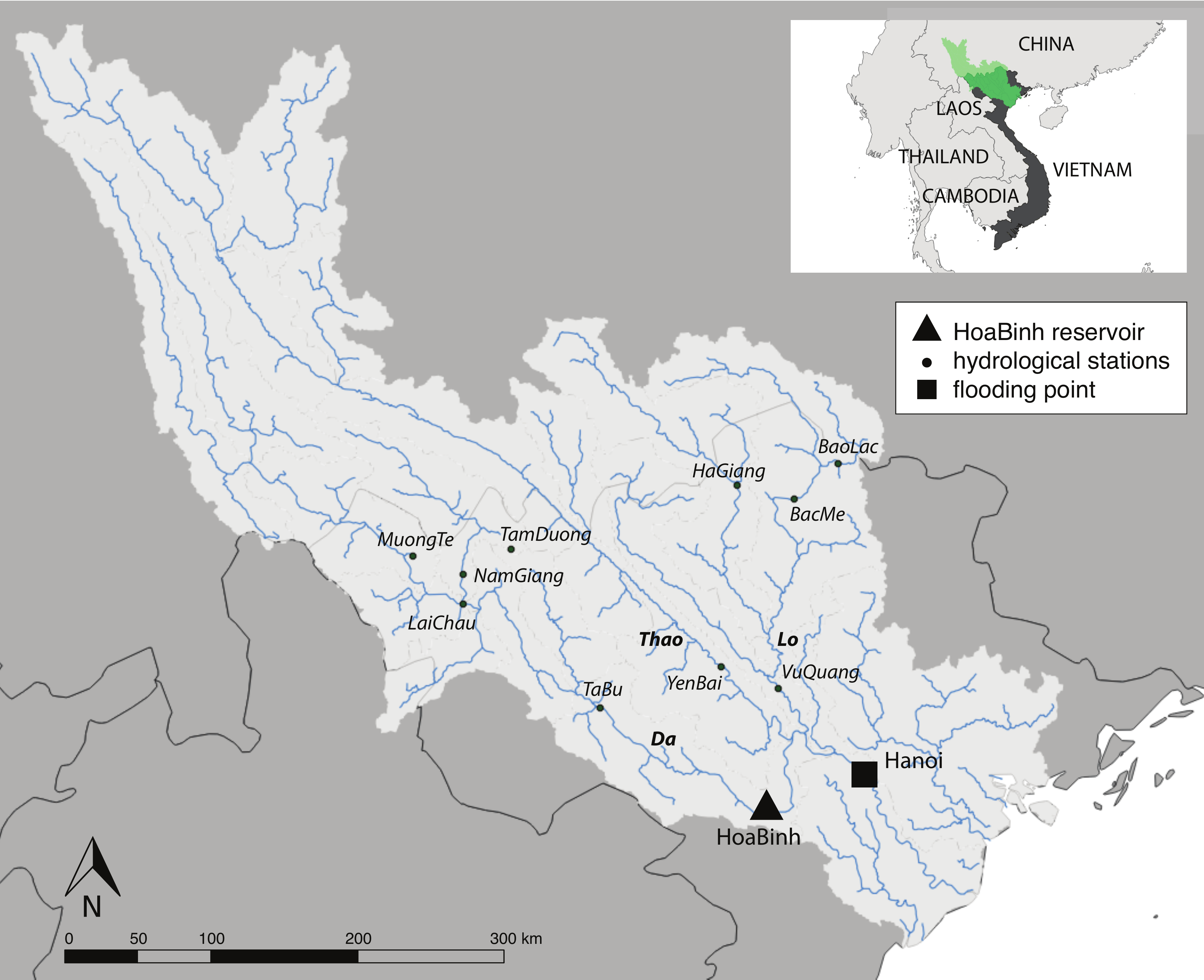}
    \caption{Map of the Red River Basin.}
    \label{fig:map}   
\end{figure}

The Hoa Binh reservoir is one of the largest reservoir in Vietnam characterized by a surface area of about 198 km$^2$ and an active storage capacity of about 6 billion m$^3$. The reservoir is located along the Da River (see Figure \ref{fig:map}), which is the main tributary of the Red River, with this latter being the second largest \new{river} basin of Vietnam accounting for a total area of about 169,000 km$^2$. The dam is connected to a power plant equipped with eight turbines, for a total design capacity of 1,920 MW, which guarantees a large share of the national electricity production. Moreover, the dam operation contributes to the control of downstream floods, particularly in the highly densely populated capital city Hanoi. 

The Hoa Binh system is here modeled as a discrete-time, periodic, non-linear, stochastic Markov Decision Process (MDP) by combining conceptual and data-driven models. More specifically, the Hoa Binh dynamics is represented by the mass balance equation of the water volume $s_t$ stored in the reservoir affected by a stochastic disturbance $q^D_{t+1}$ (namely, the inflow to the reservoir in the time interval $[t,t+1)$), i.e.
\begin{equation}\label{eq:massbalance}
    s_{t+1}=s_t+q^D_{t+1}-r_{t+1}
\end{equation}    

The nonlinear dynamics of the reservoir are due to the release function, which determines the actual release as
\begin{equation}\label{eq:release}
r_{t+1}=f(s_t,u_t,q^D_{t+1})
\end{equation} 

as a function of the control $u_t$, along with the minimum and maximum releases that can be produced in the time interval $[t, t+1)$ starting from $s_t$ with inflows $q^D_{t+1}$ (by keeping all the dam's gates completely closed and completely open, respectively).

These constraints of the problem are embedded in the model \cite{soncini2007integrated}, thus guaranteeing the feasibility of the designed solutions. In the adopted notation, the time subscript of a variable indicates the instant when its value is deterministically known. The reservoir storage $s_t$ is observed at time $t$, whereas the inflow and release have subscript $t+1$ as they depend on the realization of the stochastic process in the time interval $[t,t+1)$.
The routing of the water released from the reservoir to the delta and the city of Hanoi is simulated by a feedforward neural network providing the water level in Hanoi as a function of the Hoa Binh release along with the natural discharges of the Thao ($q^T_{t+1}$) and Lo ($q^L_{t+1}$) rivers \cite{pianosi2012identification}. 

Overall, the modelled system includes two state variables $\mathbf{x}_t = [t, s_t]$, one control $u_t$, and a vector of three stochastic disturbances $\varepsilon_{t+1} = [ q^{D}_{t+1}, q^T_{t+1},q^L_{t+1} ]$ which are both spatially and temporally correlated. 
For more details about the model formulation, see \cite{Castelletti2012}.

\begin{figure}[h]
    \centering
    \includegraphics[width=0.25\textwidth]{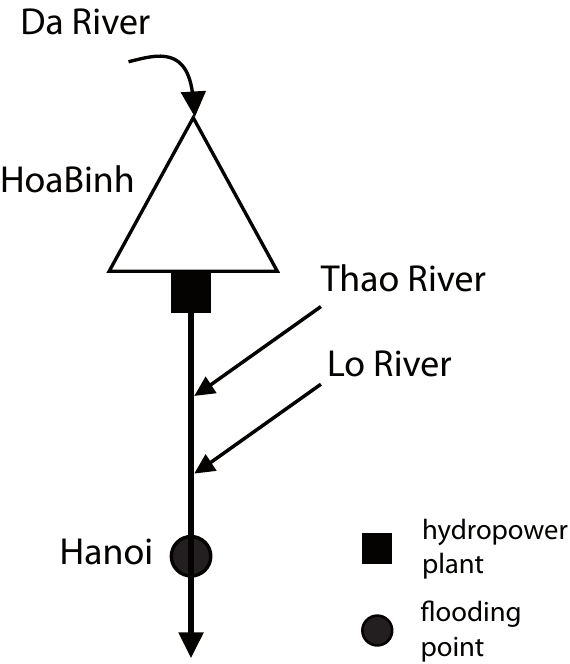}
    \caption{Schematic representation of the model.}
    \label{fig:model}
\end{figure}

\subsection{Policy design via dynamic programming}
%\textcolor{blue}{introduce the two conflicting interests and the problem, difference DDP and SDP, motivation for an online approach}

The two conflicting interests of hydropower production and flood control that drive the Hoa Binh operation are modeled using the following objective formulations, evaluated over the simulation horizon $T$:
\begin{itemize}
\item[-] \emph{hydropower production}: daily average energy production (kWh/day) to be maximized, computed as
\begin{equation}\label{eq:Jhyd}
	J^{H}=\frac{1}{T}\sum_{t=0}^{T-1} \left( \eta g \gamma_w \bar{h}_t q_{t+1}^{Turb} \right)\cdot 10^{-6}
\end{equation}
where $\eta$ is the turbine efficiency (which depends on the hydraulic head), $g=9.81$ (m/s$^2$) is the gravitational acceleration, $\gamma_w=1000$ (kg/m$^3$)  is the water density, $\bar{h}_t$ (m) is the net hydraulic head (i.e., reservoir level minus tailwater level), $q_{t+1}^{Turb}$ (m$^3$/s) represents the turbined flow;

\item[-] \emph{flood control}: the daily average excess level $h^{Hanoi}_{t+1}$(cm$^2$/day) in Hanoi with respect to the flooding threshold $\bar{h}$ = 950 cm, to be minimized, computed as

\begin{equation}\label{eq:Jflo}
	J^{F}=\frac{1}{T}\sum_{t=0}^{T-1} max( h^{Hanoi}_{t+1} - \bar{h},0 )^2
\end{equation}
where $h^{Hanoi}_{t+1}$ is the level in Hanoi estimated by the routing model.
\end{itemize} 

The optimal control problem of the Hoa Binh reservoir is formulated as designing the set of Pareto optimal (or approximate) control policies $\mathcal{P}^\ast$ that minimize the objective functions vector $\mathbf{J}$:
\begin{equation}\label{eq:HBproblem}
	\mathcal{P}^\ast = \arg \min_\mathcal{P} \mathbf{J} = \arg \min_\mathcal{P}  \vert -J^H , J^F \vert
\end{equation}
subject to the dynamics of Hoa Binh reservoir, see eq. \eqref{eq:massbalance}, and the objective functions are defined in eqs. \eqref{eq:Jhyd}-\eqref{eq:Jflo}. 
\new{Problem \eqref{eq:HBproblem} does not yield a single solution that minimizes the two competing objectives as it does not generally exist. Instead, it determines the set of Pareto optimal solutions $\mathcal{P}^\ast$, defined in Definition \ref{def_pareto_optimal}, which maps onto the so-called Pareto front. 
\begin{definition}\label{def_pareto_optimal}
With respect to Problem \eqref{eq:HBproblem}, a policy $p$ is said to \textit{dominate} policy $p'$, denoted by $p \prec p'$, if:
$$\begin{array}{c}
    \forall i \in \{ 1, \ldots, M\}, J^i(p) \leq J^i(p'), \\
    \wedge \\
    \exists i \in \{ 1, \ldots, M \}, J^i(p) < J^i(p')
\end{array}$$
If there is no policy $p'$ such that $p' \prec p$, the policy $p$ is Pareto optimal.
\end{definition}
The traditional approach to solve the multi-objective problem \eqref{eq:HBproblem} is to reformulate it as a series of single-objective problems $\mathcal{P}_i$ defined as follows:
\begin{equation}\label{eq:HBproblem_single_obj}
	\mathcal{P}_i = \arg \min_\mathcal{P} J^i = \arg \min_\mathcal{P}  -\alpha_i J^H + (1-\alpha_i) J^F,
\end{equation}
where $\alpha_i$ is a ponderation used to balance the two competing objectives. In practice, the Pareto optimal policies are obtained by solving Problem \eqref{eq:HBproblem_single_obj} for a set of ponderations $\{\alpha_i\}_{i=1}^M$ ranging between 0 and 1.}

These single-objective problems can be solved via \new{Dynamic Programming (DP) \cite{Bellman1957}} by computing the Bellman function \new{$H_t(\cdot)$ by solving recursively backwards the Bellman equation}:
\begin{equation}\label{eq:Bellman}
 H_t(s_t) = \min_{u_t} E_{\varepsilon_{t+1}} \left[ G_{t+1}(s_t, u_t, \varepsilon_{t+1}) + H_{t+1}(s_{t+1}) \right]
\end{equation}
where $H_t(\cdot)$ estimates the expected long-term cost of a policy over a discrete grid of states  (i.e., reservoir storage $s_t$ and time $t$) for the ``scalarized'' objective, and $G_{t+1}(\cdot)$ being the corresponding scalarized immediate cost function. 

\new{Once the Bellman function is computed, the} optimal control policy, defined as a periodic sequence\footnote{Namely: $p^\ast \triangleq [ \mu^\ast _0(s_t), \ldots, \mu^\ast _{T-1}(s_{T-1}) ]$.} of control laws with period $T$ that minimize a given scalar objective, is then derived as
\begin{equation}\label{eq:optpolicy}
	\mu^\ast_t(s_t) = \arg \min_{u_t} E_{\varepsilon_{t+1}} \left[ G_{t+1}(s_t, u_t, \varepsilon_{t+1}) + H_{t+1}(s_{t+1}) \right]
\end{equation}

\new{In this approach, the disturbances $\varepsilon$ can be considered to be deterministic over the considered considered horizon, in which case the resolution of the problem is referred to as Deterministic Dynamic Programming (DDP). In \cite{giuliani2016curses}, the disturbances are considered to be stochastic, which constitutes a more realistic approach. In this case, the single-objective problem are solved through Stochastic Dynamic Programming (SDP).}

\new{The Pareto fronts, obtained for the time period 1962-1969 with both DDP and SDP, are represented on Figure \ref{fig:pareto} with the daily average energy production on the y-axis and the daily average excess level of water on the x-axis. Good performances then correspond to the top-left corner of the graph. Each point of the curve correspond to a value $\alpha_i=\left[0,\ 0.05,\ 0.1,\ 0.2,0.4,0.5,0.6,0.8,0.9,0.95,1\right]$ (from left to right). In this paper, the policy corresponding to $\alpha_i=0.05$ is selected as the optimal one resulting from DDP.}

\subsection{\new{Objectives for future control strategies}}
In principle, \new{SDP} can solve problem \eqref{eq:optpolicy} under relatively mild assumptions \cite{Castelletti2012b}. 
In practice, the application of \new{SDP} in large-scale control schemes is constrained by the well-known \emph{curse of dimensionality} \cite{Bellman1957}. Besides, \new{SDP} is also constrained by the \emph{curse of modeling} \cite{Tsitsiklis1996}, as any input of the control policy must be explicitly modeled, and the \emph{curse of multiple objectives} \cite{Powell2007} as the generation of the full set of Pareto optimal solutions factorially scales with the growth in the number of the objectives. These three curses, along with the challenges related to the adaptation to variable hydrologic regimes, motivate the search for scalable and more flexible solutions. \new{To sum up, the following items should be taken into account when developing  a new control strategy:
\begin{enumerate}
    \item The considered system is \textbf{complex} because of the presence of nonlinearities in \eqref{eq:release} and of the flow routing model, given as an artificial neural network as developed in \cite{pianosi2012identification};
    \item The considered system is \textbf{partially unmodelled} but historical data is available;% and should be taken advantage of when designing a controller.
    \item The \textbf{evolution of the hydrologic conditions} should be taken into account. Indeed, it may reduce the performances of the optimal policy $p^\star$ found offline through SDP for a given time-period when applied to another time period \cite{giuliani2016large};
    \item Two \textbf{conflicting objectives} are considered, the power production in \eqref{eq:Jhyd} to be maximized and the flood damages in \eqref{eq:Jflo} to be minimized.
\end{enumerate}
To overcome these challenges, the strategy proposed in this paper, detailed in the next section, combines data-driven control, to tackle items 1) and 2), and an online model-based controller in order to handle items 3) and 4). In the rest of this paper, the performances of the proposed control strategy will be compared with the policy obtained through SDP in \cite{giuliani2016curses}. Meanwhile, the policy obtained through DDP is to be seen as the optimal behaviour on the considered time period and, while this strategy cannot be realistically be implemented, its results should be regarded as the ideal case one wants to get closer to. Before moving on to the proposed control strategy, it is worth recalling that the proposed design relies on different types of data, listed hereafter:
\begin{itemize}
    \item \textbf{Historical data} from the Hoa Binh reservoir: the daily water inflows from the Da, Thao and Lo rivers are known from 1959 to 2008. In this paper, two time periods are considered. The first one is 1962-1969, as selected in \cite{giuliani2016curses} as it comprises normal, wet, and dry years. It is used as a training data-set for the proposed approach, while the second period is 2007-2008 and is used as a verification set for the proposed control strategy to test the different approaches under different hydroclimatic conditions;
    \item The \textbf{release function} $f$ is not parametrized: the minimal and maximal release are known only on a grid of values of reservoir storage and inflows from the Da river;
    \item The \textbf{optimal policy} obtained through DDP with $\alpha_i=0.05$ for the first time period (1962-1969) is denoted $\{u^\star, s^\star\}$ and embeds the historical data from the same period. It allows to compute the ideal mean annual behaviour $\{u^c, s^c\}$, which is used in the proposed approach to learn a low-level controller through data-driven design.
\end{itemize}
}

\section{Hierarchical control system design}
\label{sec:control_design}
\new{This paper aims at proposing an online strategy, such as economic Model Predictive Control (eMPC), to operate such water reservoir systems since it would enable to better handle uncertainty, as the evolution of the hydrologic inputs could be taken into account. This strategy would also allow to enforce path constraints as in the classical SDP approach, ensuring that the obtained policy is feasible. However, the complexity and non-linearity of water reservoir systems constitutes a major obstacle to the implementation of such strategy as the online resolution of the associated optimization problem would represent a significant computational burden.}

\begin{figure}
    \centering
    \scalebox{0.85}{\tikzstyle{abstract}=[rectangle, draw=black, rounded corners, fill=blue!5, text centered, anchor=north, text=black, text width=3.5cm]
\tikzstyle{abstract2}=[rectangle, text centered, anchor=north, text=black]
\tikzstyle{block}=[rectangle, draw=black, text centered]
\tikzstyle{comment}=[rectangle, draw=black, rounded corners, text centered, anchor=north, text=black, text width=2cm]
\tikzstyle{myarrow}=[->, >=open triangle 90, thick]
\tikzstyle{line}=[-, thick]
\tikzstyle{sum}=[draw,circle,text width = 0.3cm]

\begin{tikzpicture}
	\draw
    
	node at (-0.5,0)[block, fill=red!10](MPC){eMPC}
	node at (-0.5,1.5)[comment](predictions){Prediction of
	$q^{D}, q^T,q^L$}
	node at (1.5,0)[sum](comp){}
	node at (2.5,0)[block, fill=red!10](cor){$C$}
	node at (4,0)[block](NLf){$f$}
	node at (5.5,0)[sum](plant){}
	node at (5.5,1.5)[](qd){}
	node at (7,0)[](y1){};

	\draw (comp.135) -- (comp.315);
    \draw (comp.225) -- (comp.45);
    \node (plus) [below] at (comp.180) {$+$};
    \node (minus) [below] at (comp.350) {$-$};
    
    \draw (plant.135) -- (plant.315);
    \draw (plant.225) -- (plant.45);
    \node (minus2) [below] at (plant.180) {$-$};
    \node (plus2) [above] at (plant.20) {$+$};
	
	\draw[->](predictions) -- (MPC);
	\draw[->](MPC) -- node[above]{$s^{ref}$}(comp);
	\draw[->](comp) -- node[above]{}(cor) -- node[above]{$u$}(NLf) -- node[above]{$r$}(plant) --node[above]{$s$}(y1);
	\draw[->](6,0) -- (6,-1) -| node[below]{}(comp);
    \draw[->](qd) -- node[right]{$q^D$}(plant);
    \draw[-, dash pattern=on 2pt off 1pt, line width=0.5mm, red!50] (0.9,0.6) -- (6.5,0.6) -- (6.5,-1.5) -- node[below]{Inner loop}(0.9,-1.5) -- (0.9,0.6) ;
\end{tikzpicture}}    
    \caption{The proposed control architecture: the inner-loop is designed to simplify the model dynamics for the outer reference governor, while the latter consists of an economic model predictive controller generating $s^{ref}$.}
    \label{fig:control_architecture}
\end{figure}
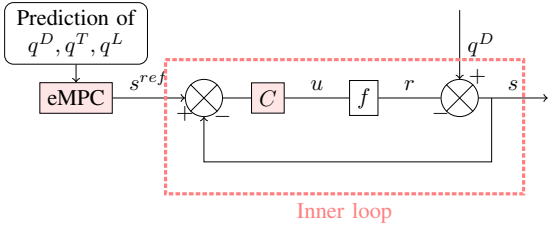

\new{To that extent, the hierarchical approach introduced in \cite{piga2017direct} allows to overcome this issue while still offering the benefits of an online strategy. Indeed,} the first step consists in reducing the complexity of the system dynamics by closing an inner loop through a model-free feedback controller: the design of such a low-level controller $C$ allows to assimilate the inner-loop to a simpler linear model\new{, which can be used as a prediction model in an outer loop eMPC.} This whole hierarchical approach is detailed in this section, starting with the design of the inner-loop controller in \ref{subsec:inner_loop}. The overall control architecture is shown in Figure \ref{fig:control_architecture}.

\subsection{Data-driven inner-loop design}
\label{subsec:inner_loop}
To start with, a low-level controller $\mathbf{C}$ is designed to reduce the complexity of the system's behaviour and to transform it into a known linear inner-loop model. The system is described by the balance equation \eqref{eq:massbalance} and by the nonlinear release function $f$. Since $f$ does not have a parametrized expression, it motivates the use of a data-driven technique, such as the VRFT \cite{campi2002virtual}, to design the inner-loop controller. 

\begin{figure}[h]
    \centering
    \scalebox{0.85}{\tikzstyle{abstract}=[rectangle, draw=black, rounded corners, fill=blue!5, text centered, anchor=north, text=black, text width=3.5cm]
\tikzstyle{abstract2}=[rectangle, text centered, anchor=north, text=black]
\tikzstyle{block}=[rectangle, draw=black, text centered]
\tikzstyle{comment}=[rectangle, draw=black, rounded corners, fill=red!5, text centered, anchor=north, text=black, text width=3cm]
\tikzstyle{myarrow}=[->, >=open triangle 90, thick]
\tikzstyle{line}=[-, thick]
\tikzstyle{sum}=[draw,circle,text width = 0.3cm]

\begin{tikzpicture}
	\draw
	
	node at (0,0)[]{}
	node [name=ref] {} 
	node at (1.5,0)[sum](comp){}
	node at (3,0)[block, fill=red!10](cor){$C(z,\theta)$}
	node at (3,1)[block](cor2){$M(z)$}
	node at (5,0)[block](plant){$P(z)$}
	node at (7,0)[](y1){}
	node at (5,1)[](y2){};

	\draw (comp.135) -- (comp.315);
    \draw (comp.225) -- (comp.45);
    \node (plus) [below] at (comp.180) {$+$};
    \node (minus) [below] at (comp.350) {$-$};
	
	\draw[->](ref) -- node[above]{$\overline{r}$}(comp);
	\draw[->](comp) -- node[above]{$\overline{e}$}(cor) -- node[above]{$u^c$}(plant) --node[above]{$s^c$}(y1);
	\draw[->](1,0) |- (cor2) -- node[above]{$s^c$}(y2);
	\draw[->](6,0) -- (6,-1) -| node[below]{}(comp);
	\draw[-, dash pattern=on 2pt off 1pt, line width=0.5mm, red!50] (0.8,0.6) -- (6.5,0.6) -- (6.5,-1.5) -- node[below]{Inner loop}(0.8,-1.5) -- (0.8,0.6) ;
	\end{tikzpicture}}
    \caption{Inner-loop design using VRFT: $\mathbf{M}$ is the reference model, $\mathbf{P}$ is the system to be controlled and $\mathbf{C}$ represents the controller to be tuned on the basis of the signals $u^c$ and $s^c$. The signals $\overline{r}$ and $\overline{e}$ are the virtual reference and error respectively.}
    \label{fig:inner_loop_design}
\end{figure}
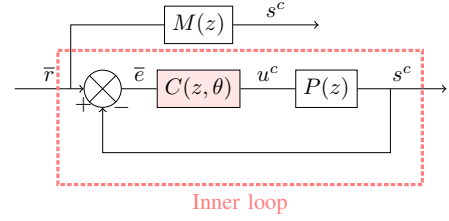

Based on the available time-domain data $\{u^c,s^c\}$, a desired closed-loop behaviour $M$ and a controller structure $C(\theta)$, the objective of the VRFT is to find the controller parameters $\theta^\star$ such that the resulting closed-loop is as close as possible to the reference model $M$. The key idea is the computation of the virtual reference signal
\begin{equation}
\overline{r}_t = M^{-1}(z)s^c_t, 
\end{equation}
as the reference that would feed the loop if the complementary sensitivity function was exactly $M$. Like that, the optimal controller (i.e., the one achieving $M$ in closed-loop) can be computed as the system producing $u^c_t$ when fed by the virtual error $\overline{e}_t=\overline{r}_t-s^c_t$. Since the virtual error and the virtual reference can be computed off-line based on the available dataset, also the controller is retrieved by solving the \emph{one-shot} optimization problem
\begin{equation}
    \theta^\star = \underset{\theta}{\textnormal{arg min}} \frac{1}{N}\sum_{t=1}^N (u^c_t-\mathbf{C}(z,\theta)\overline{e}_t)^2.
    \label{VRFT_pb}
\end{equation}
Specifically, in \cite{campi2002virtual}, it has been shown that a linear parameterization of the controller $C(\theta)$ (like a PID) leads to a simple quadratic problem that can be solved via least squares formulas. By suitably prefiltering $u_t$ and $s_t$, it can also be shown that the minimum of \eqref{VRFT_pb} coincides with the optimal controller, if this belongs to the considered class \cite{formentin2019deterministic}.

In this work, the cyclostationnary signals $u^c$ (release decision) and $s^c$ (storage), which represents the annual mean behaviour \new{obtained from the DDP ideal policy for 1962-1969}, are used as the dataset for the identification of the controller. The desired closed-loop behavior $M$ is defined as
\begin{equation}
    M(z)=\frac{0.2z^{-1}}{1-0.8z^{-1}}.
    \label{eq:ref_model}
\end{equation}

Unlike the classical VRFT framework, here the reference model does not need to represent the desired closed-loop performance: this will be handled by the outer-loop eMPC controller. The reference model only needs to be practically achievable, which is most likely the case when specifying a low-performance closed-loop behavior $M$.

In the present case, the inner-loop controller has a PID structure:
\begin{equation}
    C(z,\theta)=\theta_1+\frac{\theta_2}{1-z^{-1}}+\theta_3(1-z^{-1})
    \label{eq:struct_C}
\end{equation}
and the coefficients obtained through the VRFT procedure are 
\begin{equation}
    \theta^\star=10^{-6}\left[-0.4439 \ -0.1063 \ -0.1898\right]^T.
\end{equation}

The resulting inner-loop, see Figure \ref{fig:control_architecture}, contains a nonlinear block $f$ which corresponds to the physical limitations of the dam in terms of water release. 

\begin{figure*}
    \centering
    \subfloat{\includegraphics[width=0.48\textwidth]{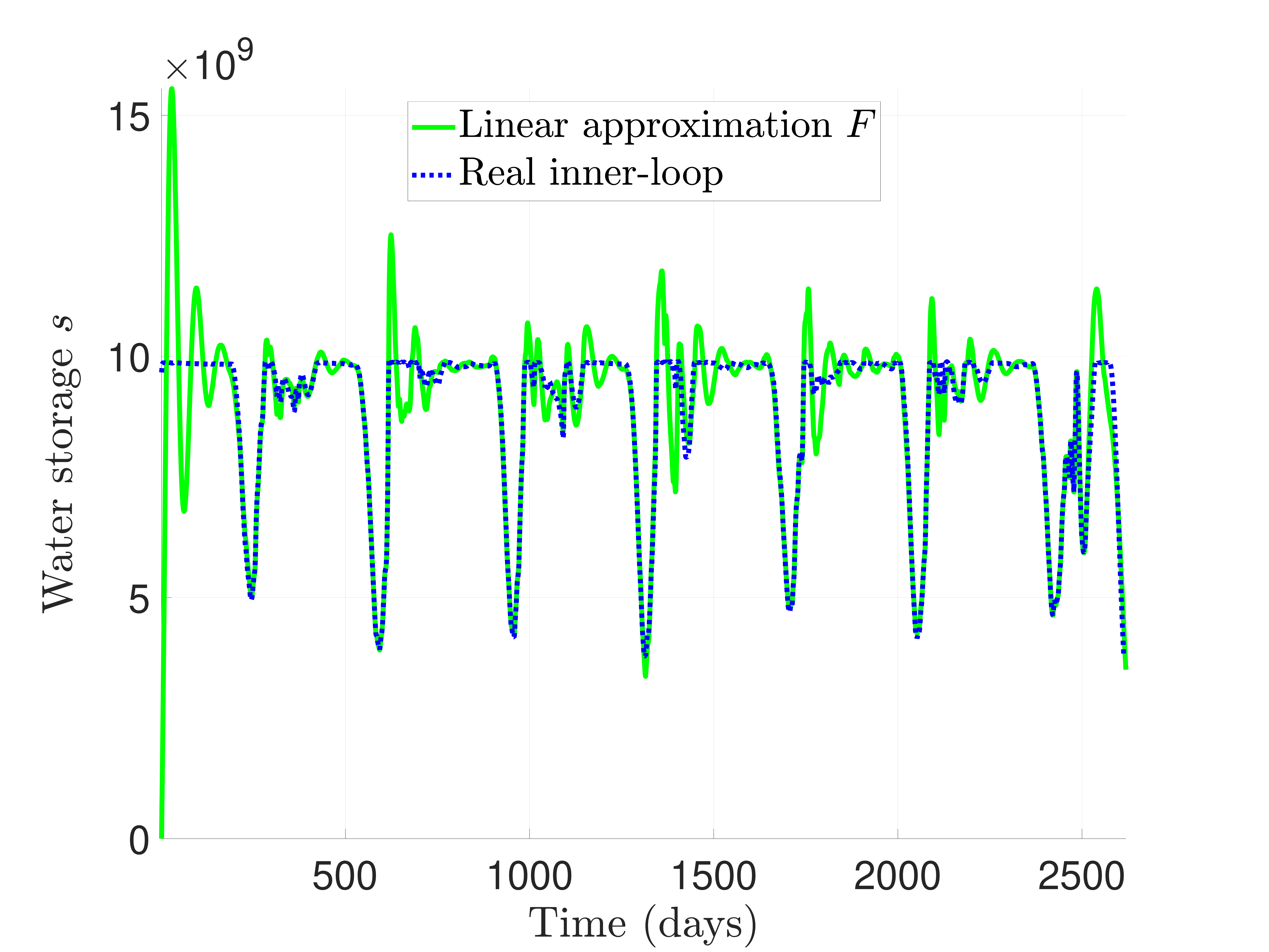}\label{inner_loop_water_storage}}
    \subfloat{\includegraphics[width=0.48\textwidth]{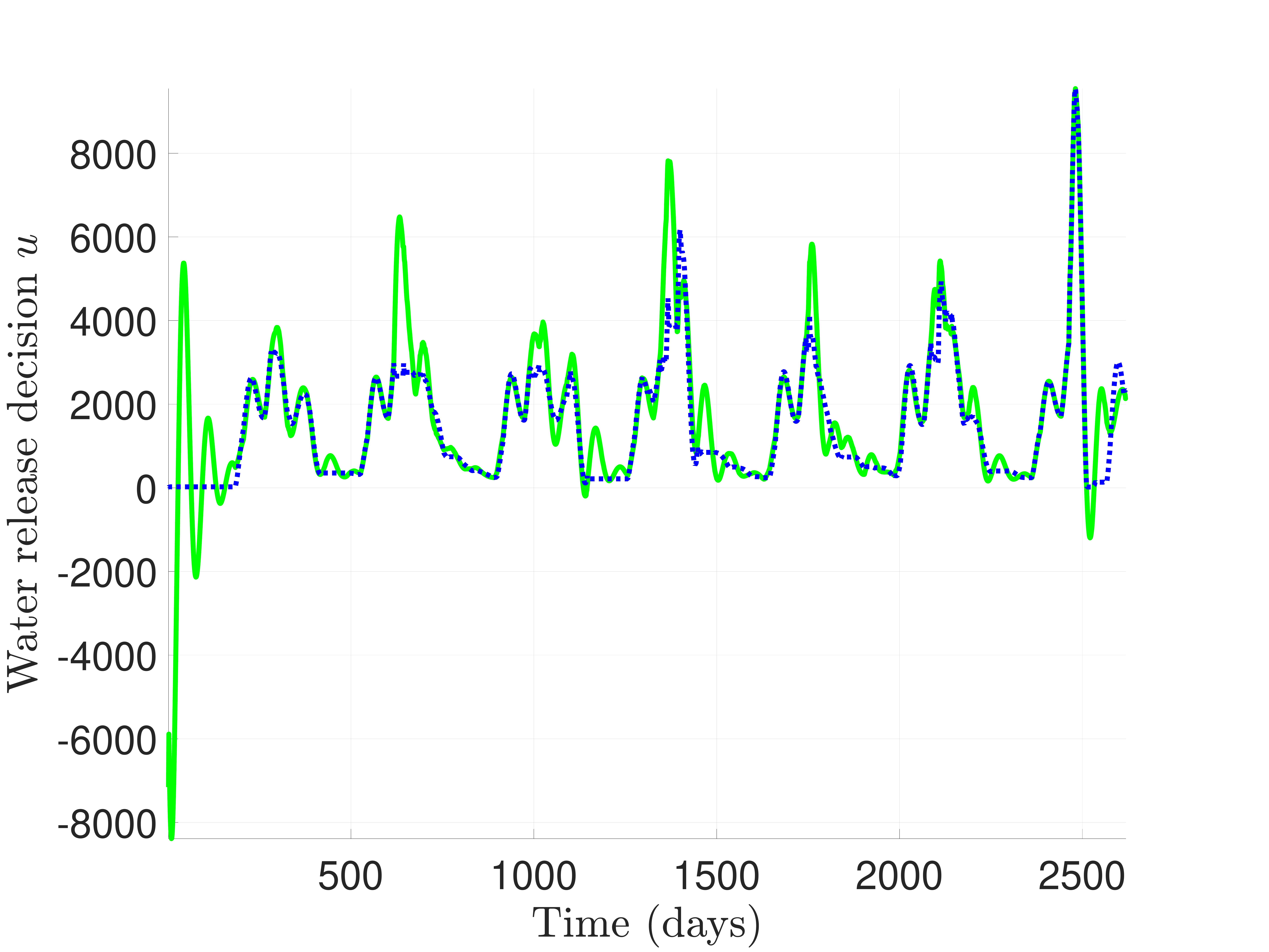}\label{inner_loop_water_release}}
    \caption{Performance of the inner-loop and linear approximation $F$ of the inner-loop, see eq. \eqref{eq:linear_approximation}. The water release decision $u$ (left) and water storage $s$ (right) are obtained by simulating the true inner-loop and its linear approximation  over the period 1962-1969. The signal $s^{ref}$ is taken as a desirable water storage signal, obtained by solving the optimal control problem through DDP over the same period.}
    \label{fig:linear_inner_loop}
\end{figure*}

The purpose of the inner-loop controller $C$ is to assimilate the inner-loop to a simple linear model, which will be used later for prediction in the outer-loop to simplify the eMPC problem. To that extent, the nonlinear term $f$ is neglected, i.e. $u\approx r$, which leads to the linear model $F$ approximating the inner-loop behaviour:
\begin{equation}
    F: \ \left\{\begin{array}{r@{=}l}
    s(z) & z^{-1}s(z)+T_s(q^D(z)-z^{-1}u(z))  \\
    u(z) & C(z)(s^{ref}(z)-s(z))
\end{array} \right. .
\label{eq:linear_approximation}
\end{equation}

Figure \ref{fig:linear_inner_loop} compares the water-storage signals that are obtained using the inner-loop linear approximation $F$ or by simulating the real inner-loop visible on Figure \ref{fig:control_architecture}, both fed with the optimal water storage $s^\star$ as reference signal $s^{ref}$. The signal $s^\star$ is found through DDP when solving the optimal control problem for the period 1962-1969. The results indicate that the linear model $F$ can reasonably be used to predict the inner-loop behaviour. Even though the prediction model is not perfect because of the nonlinear part $f$ of the true inner-loop, this aspect will be handled in the outer-loop through constraints enforcement. 

Finally, this simulation highlights the fact that the inner-loop alone is not sufficient for this application, for two reasons. First, \new{the choice of a practically achievable and low performance reference model $M$ as in \eqref{eq:ref_model} implies that the resulting inner-loop} may be too slow to produce a sufficient level of benefit in terms of hydropower production and floodings' damages. Besides that, the reference $s^{ref}$ to be tracked still needs to be defined. The optimal control strategy could not be used in this case, since it would require to optimize over the whole considered period. The best option is therefore to perform an online optimization of the reference signal on a limited prediction horizon. Based on these considerations and on the simple description of the inner-loop obtained through this data-driven control design, an outer-loop model predictive controller is then designed as described in the next paragraph. 

%Figure \ref{fig:linear_inner_loop} shows that the inner-loop exhibits reasonable tracking performances when fed with $s^\star$ as reference, but

\subsection{Model-based outer-loop design using economic MPC}
While the inner-loop allows to have a simplified description of the problem, the outer-loop has a double objective: performance optimization and constraints enforcement. To that extent, an economic model predictive controller is implemented, solving the following problem at each time-step $t$: 
\begin{equation}
    \begin{array}{cl}
\underset{s^{ref}}{\textnormal{min}} &  -\alpha_{eMPC}J^{hyd}_t+(1-\alpha_{eMPC})J^{flo}_t  \\
& \\
\textnormal{s.t.} & \forall k=1\dots N_p 
\end{array}
\label{eMPC}
\end{equation}
$$
\left\{\begin{array}{c@{=}c}
    x(t+k+1) & A_Fx(t+k)+ B_F\left(\begin{array}{c} s^{ref}(t+k) \\ q^D(t+k) \end{array}\right) \\
    \left(\begin{array}{c} s(t+k+1) \\ u(t+k+1) \end{array}\right) & C_Fx(t+k)+  D_F\left(\begin{array}{c} s^{ref}(t+k) \\ q^D(t+k) \end{array}\right)
\end{array}\right.
$$
$$
\color{black}{r^{min}(s_{t\new{+k}},q^D_{t\new{+k}+1})\leq u(t\new{+k})\leq r^{max}(s_{t\new{+k}},q^D_{t\new{+k}+1})}
$$
$$
\color{black}{s^{min}\leq s(t\new{+k})\leq s^{max}}
$$
\begin{figure*}
    \centering
    \subfloat[][Water release decision and actual water release.]{\includegraphics[width=0.48\textwidth]{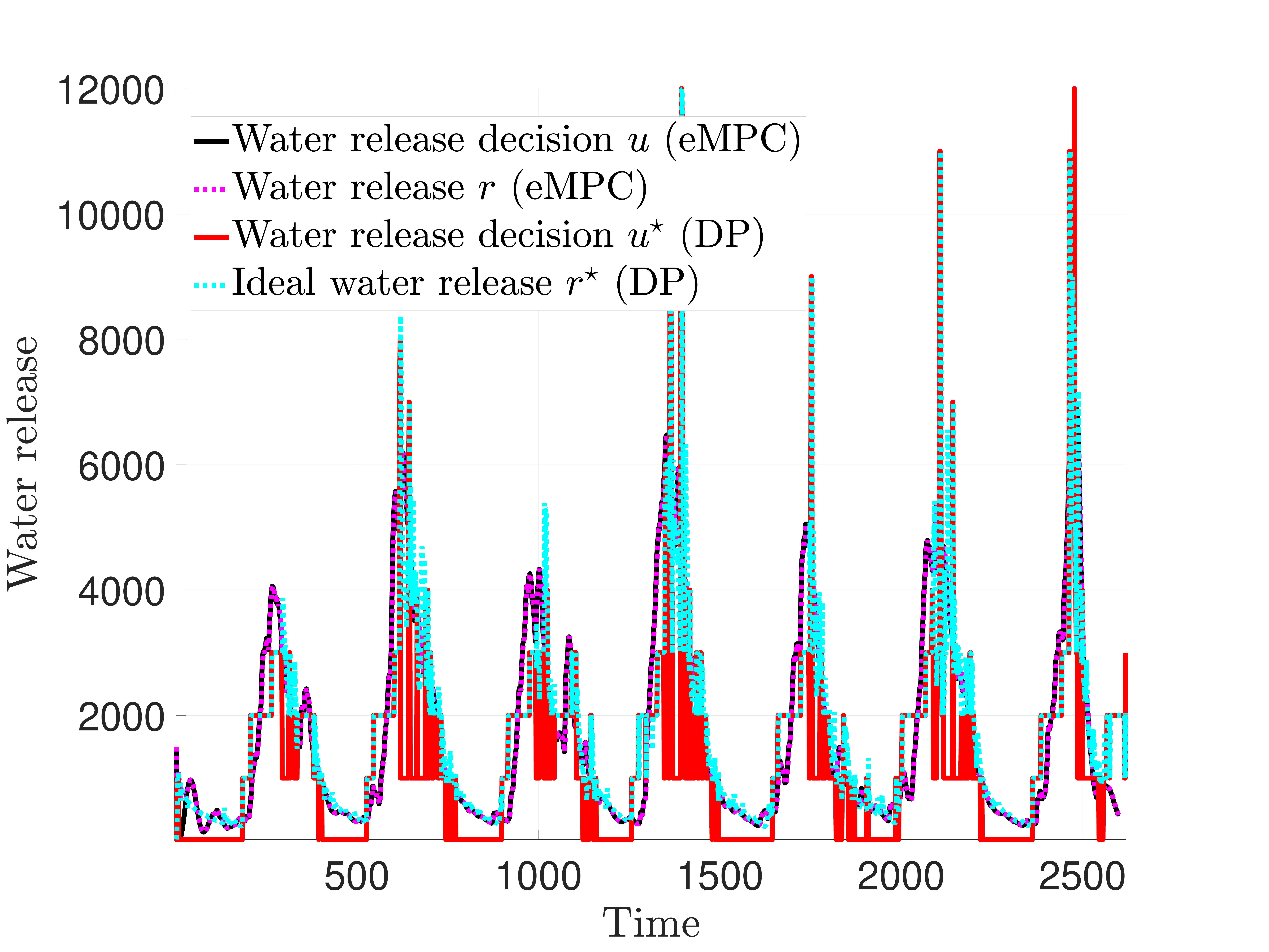}\label{u_empc}}
    \subfloat[][Water storage in the Hoa Binh reservoir.]{\includegraphics[width=0.48\textwidth]{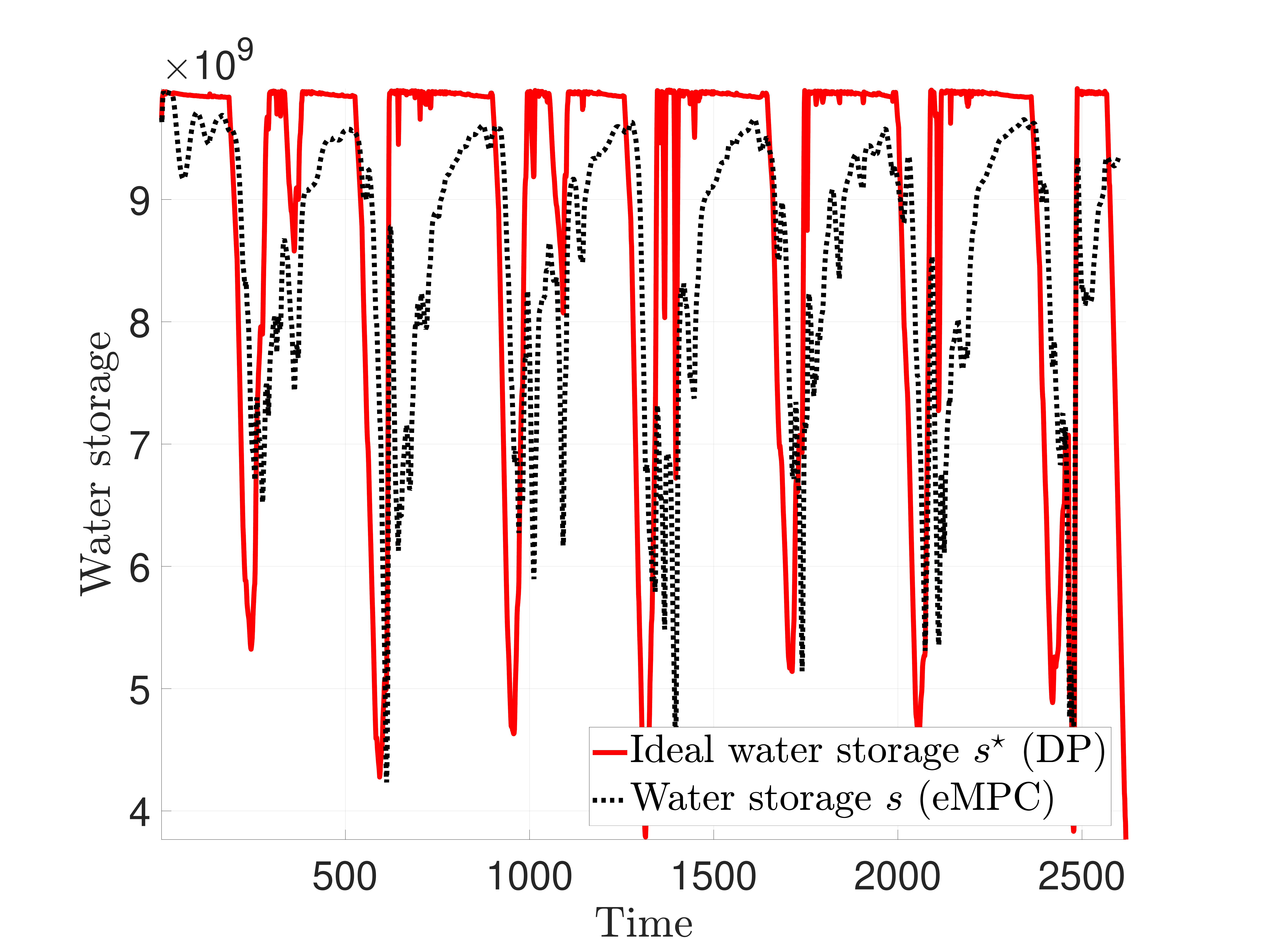}\label{s_empc}}
    \caption{Trajectory of the controlled system when combining the inner-loop controller with the outer-loop economic MPC. The controlled system is simulated over the period 1962-1969 and compared with the DDP approach.}
    \label{fig:e_mpc_trajectory}
\end{figure*}
\begin{figure*}
    \centering
    \subfloat[][Daily average energy production.]{\includegraphics[width=0.48\textwidth]{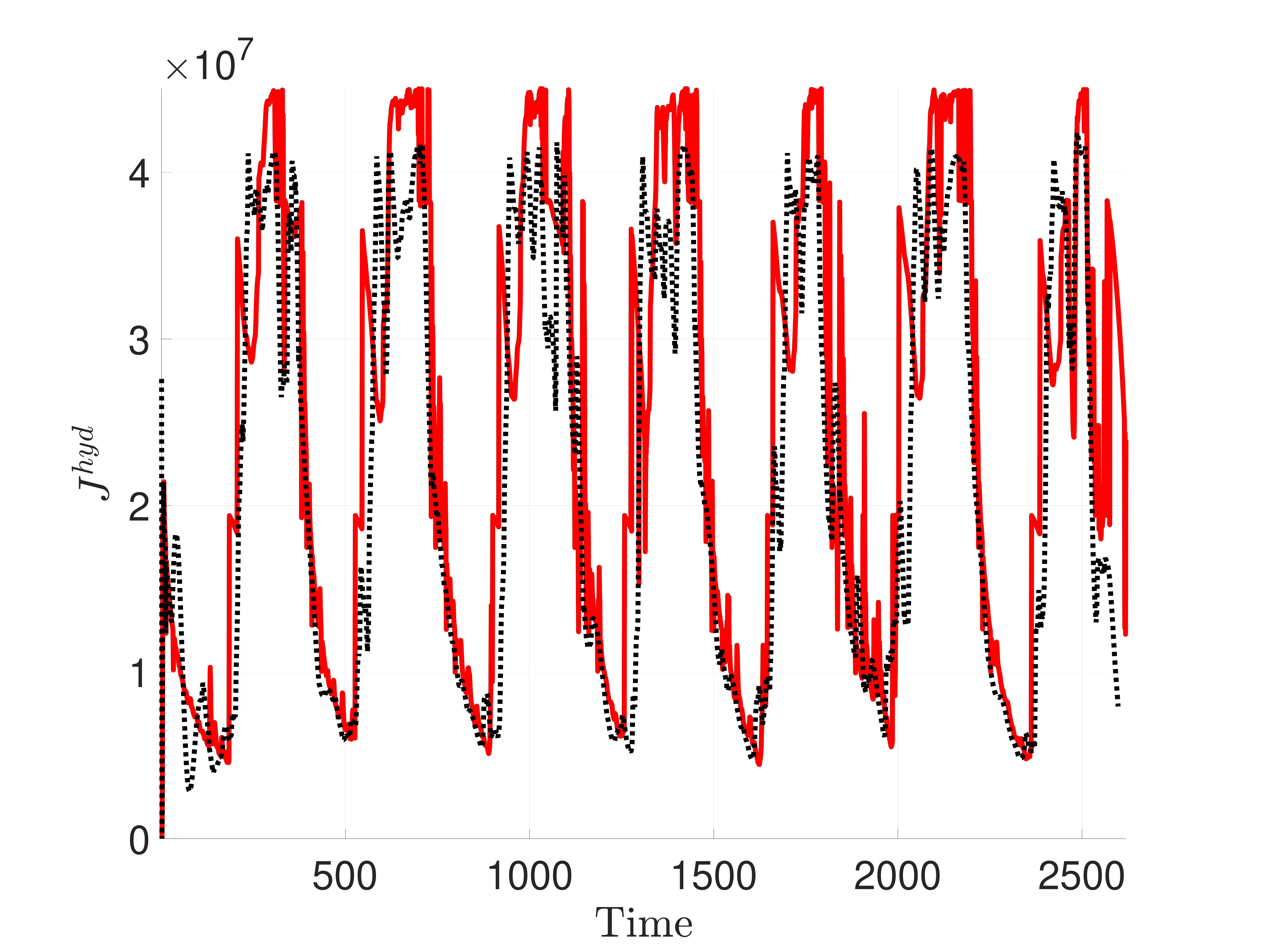}\label{Jhyd_empc}}
    \subfloat[][Daily average excess of water in Hanoi.]{\includegraphics[width=0.48\textwidth]{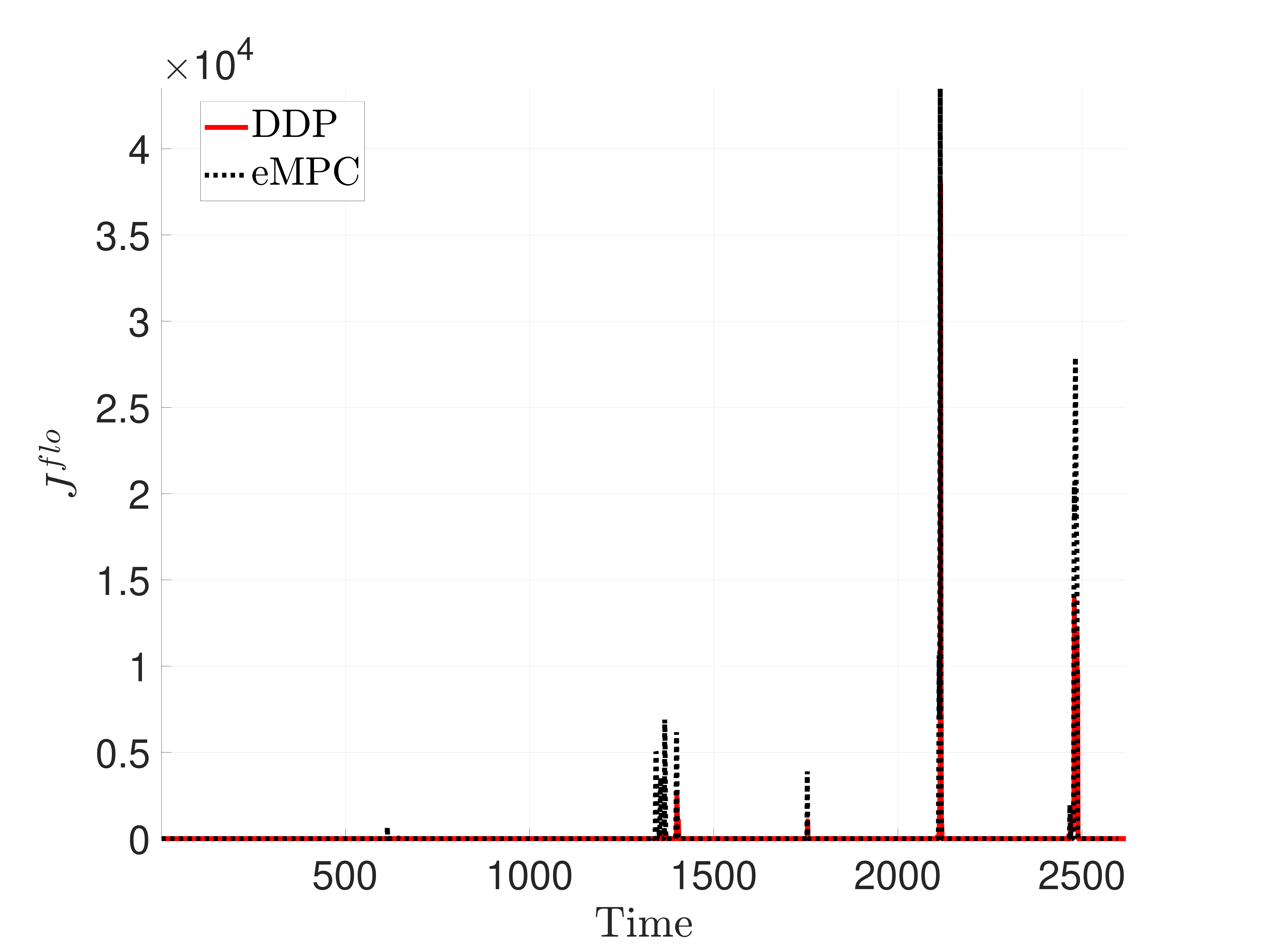}\label{Jflo_empc}}
    \caption{Performance of the controlled system in terms of hydropower production and flood damages over the period 1962-1969, compared with the DDP approach.}
    \label{fig:e_mpc_performances}
\end{figure*}
First, the outer-loop controller acts as a reference governor for the inner-loop, determining the signal $s^{ref}$ such that good performance is obtained. The performance is determined through the daily average energy production $J^{hyd}$ and the daily average excess level in Hanoi $J^{flo}$ over the prediction horizon $N_p$. Their definition is therefore similar to the ones given in \eqref{eq:Jhyd} and \eqref{eq:Jflo} for the optimal control approach.
\new{
\begin{equation}\label{eq:Jhyd_eMPC}
	J^{hyd}_t=\frac{1}{N_p}\sum_{k=t}^{N_p} \left( \eta g \gamma_w \bar{h}_k q_{k+1}^{Turb} \right)\cdot 10^{-6}
\end{equation}
\begin{equation}\label{eq:Jflo_eMPC}
	J^{flo}_t=\frac{1}{N_p}\sum_{t=0}^{N_p-1} max( h^{Hanoi}_{t+1} - \bar{h},0 )^2
\end{equation}
}
The major difference \new{with respect to \eqref{eq:Jhyd} and \eqref{eq:Jflo}} is that the performance is here evaluated online on a shorter time window, starting at the current time step $t$ and covering the prediction horizon $N_p$, instead of computing the optimal solution offline using the whole period information (in the present case the 1962-1969 period as in \cite{giuliani2016curses}). \new{As in the optimal control approach in \eqref{eq:HBproblem_single_obj}, a ponderation $\alpha_{eMPC}\in [0,1]$ is used to balance the two conflicting objectives.}

To evaluate the performances over the prediction horizon, the linear description $F$ of the inner-loop, given in a state-space form $\left(A_F, B_F, C_F, D_F\right)$, is used as prediction model (suitably scaled to avoid numerical errors). Its input are the reference water storage $s^{ref}$, which is the optimization variable, and the flow $q^D$ from the Da river. The outputs of the inner-loop are the water storage $s$ and the water release decision $u$. 

Secondly, the outer-loop controller should enforce constraints on the release decision $u$ so that the assumption $u\approx r$ made earlier holds. In addition, enforcing the constraints allow to avoid emptying the reservoir, which would happen otherwise since the flood damage objective is mostly equal to zero over time. To that extent, hard constraints are imposed on both outputs. The constraints are constant when it comes to the water storage $s$, with $s_{min}=3.8Gm^3$ and $s_{max}=9.9Gm^3$. The constraints on $u$ are nonlinear and allow to take into account the nonlinearity $f$ of the system: $r_{min}$ and $r_{max}$ represent the minimal and maximal water release, respectively. Like $f$, they depend on the current water storage $s_t$ and the incoming flow $q^D$. These functions $r_{min}$ and $r_{max}$ are not parameterized: their value is known only for a given set of values of storage and flow. \new{They are computed online according to the 2 nearest neigbors of the considered storage value and inflow from the Da river.}

\section{Simulation results}
\label{sec:results}
\begin{figure}[h]
    \centering
    \includegraphics[width=0.5\textwidth]{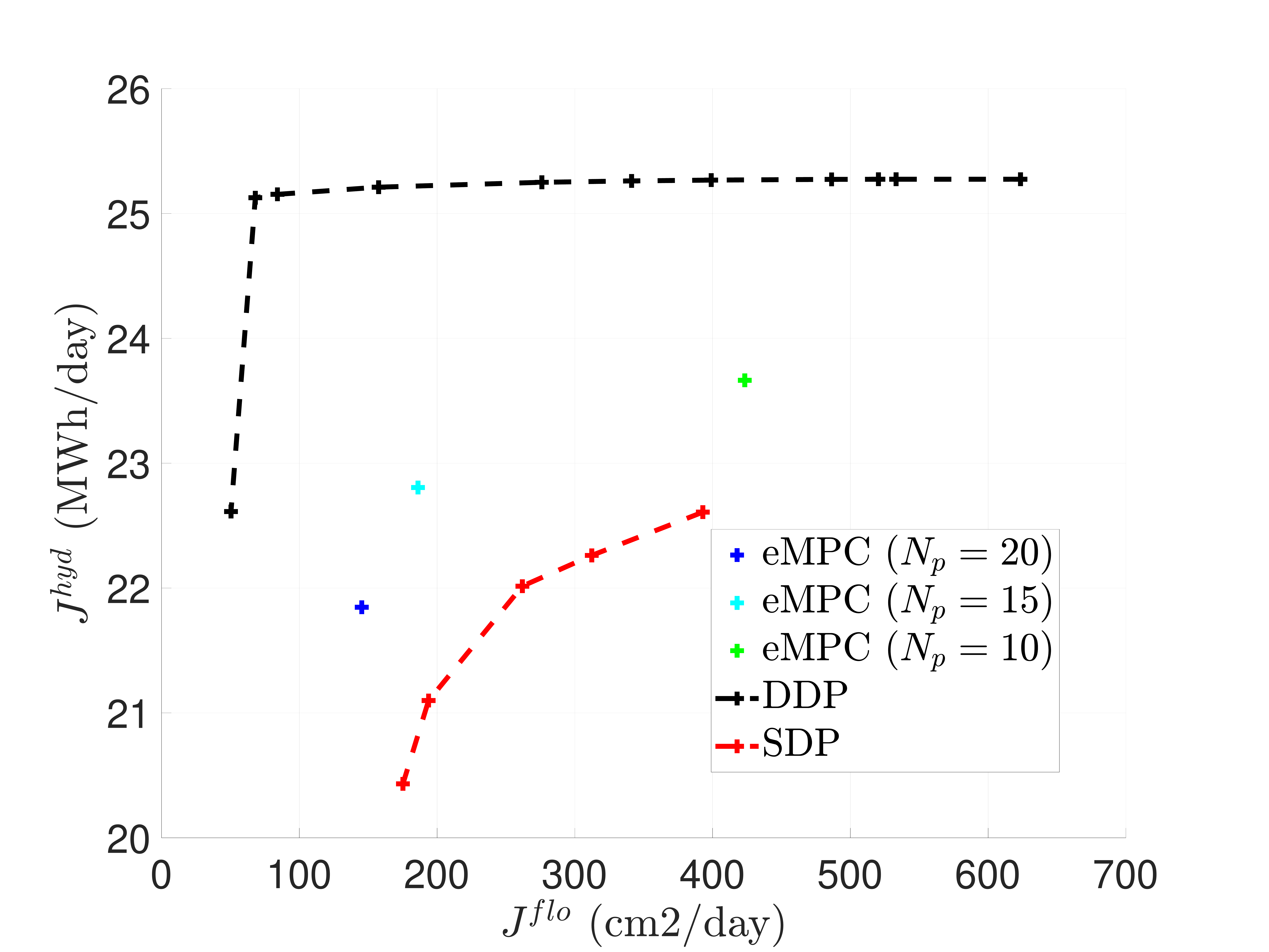}
    \caption{Pareto front representing the performance trade-off between hydropower production and excess water level in Hanoi \new{(1962-1969)} for different policies: DDP, SDP and the proposed strategy.}
    \label{fig:pareto}
\end{figure}

The behaviour of the controlled system, including both the inner-loop and the outer loop controllers, is simulated over the period 1962-1969, using the available data regarding the incoming flows $q^D$, $q^T$ and $q^L$. The prediction horizon is $N_p=15$ days and \new{the ponderation in the cost function is selected as $\alpha_{eMPC}=0.05$ since this value was chosen on the DDP Pareto front Figure \ref{fig:pareto} as the one giving the best compromise between the conflicting objectives.} 

The resulting water release and water storage are visible in Figure \ref{fig:e_mpc_trajectory}. Panel a. highlights that the constraints on the water release decision are satisfied so that $u_t=r_{t+1}$. The corresponding performance can be evaluated through the hydropower production and the excess amount of water in Hanoi, visible in Figure \ref{fig:e_mpc_performances}. The results of the optimal control approach, solved through DDP over the same period, are also represented in both the figures and can be seen as an ``upper'' reference. In comparison, the proposed control strategy exhibits a slower reaction to the monsoon, \new{mainly when refilling the reservoir every year, and does not reach the highest storage level of the DDP policy for the rest of the time.}

The performance of the controlled system can be more easily appreciated when looking at the Pareto front given in Figure \ref{fig:pareto}, which represents the compromise between the average daily hydropower production and the average daily excess of water in Hanoi. These values are computed over the simulation period 1962-1969 for three different values of the prediction horizon, $N_p=10$, $15$ and $20$ days. Shortening the prediction horizon allows to increase the hydropower production, also implies to increase the flooding in Hanoi. Indeed, a prediction horizon of 10 days is not sufficient to \new{cover the time scale of the floodings, and makes it harder to anticipate this phenomenon by adjusting the release.}

The performance compromise is also represented for the optimal control approaches DDP and SDP. The DDP approach can be conceived as an upper bound for the achievable performance.  Unlike the proposed approach, the optimization is done \emph{offline} for the whole considered period. According to Figure \ref{fig:pareto}, the proposed approach performs better than SDP, which makes it a good candidate for its online implementation and use in different time periods.%Obviously, the performance depends on the weights $\alpha^{hyd}$ and $\alpha^{flo}$ used in the cost function.  
\new{
\begin{remark}
\label{remark_flooding}
The Pareto front of the DDP and SDP approaches correspond to different values of the ponderation $\alpha_i$ in \eqref{eq:HBproblem_single_obj} while, the proposed hierarchical approach, the tradeoff varies with the prediction horizons $N_p$. Indeed, in the proposed strategy, the ponderation $\alpha_{eMPC}$ does not influence much the performance tradeoff for a given prediction horizon $N_p$. This is due to the fact that the excess level of water is mostly equal to zero by definition \eqref{eq:Jflo_eMPC}. In the future, it would be interesting to define another objective function for flood minimization purposes that would be more adequate on a short prediction horizon. 
\end{remark}
}

\begin{figure*}
    \centering
    \subfloat[][Daily average energy production.]{\includegraphics[width=0.48\textwidth]{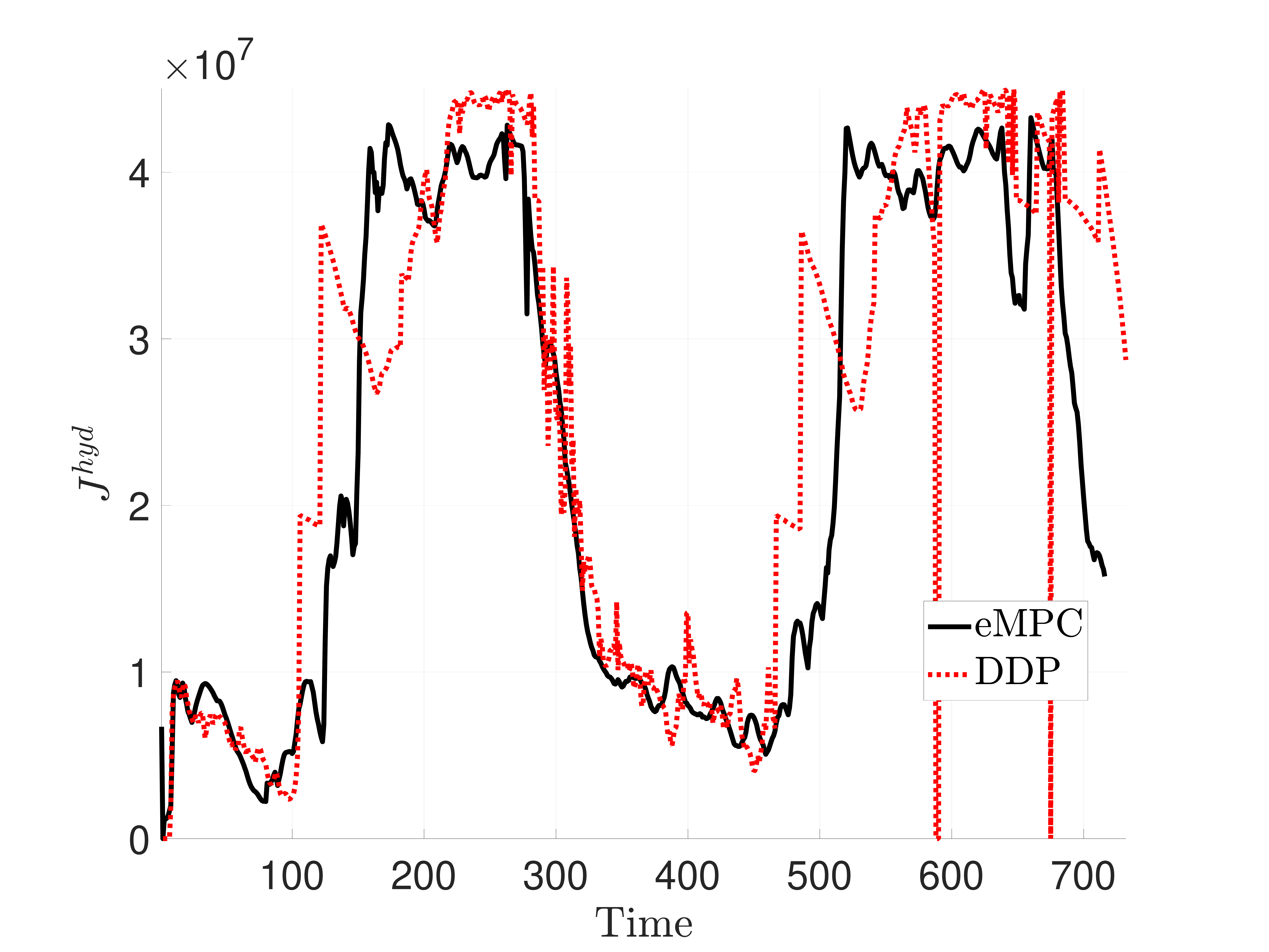}\label{Jhyd_empc_2007_2008}}
    \subfloat[][Daily average excess of water in Hanoi.]{\includegraphics[width=0.48\textwidth]{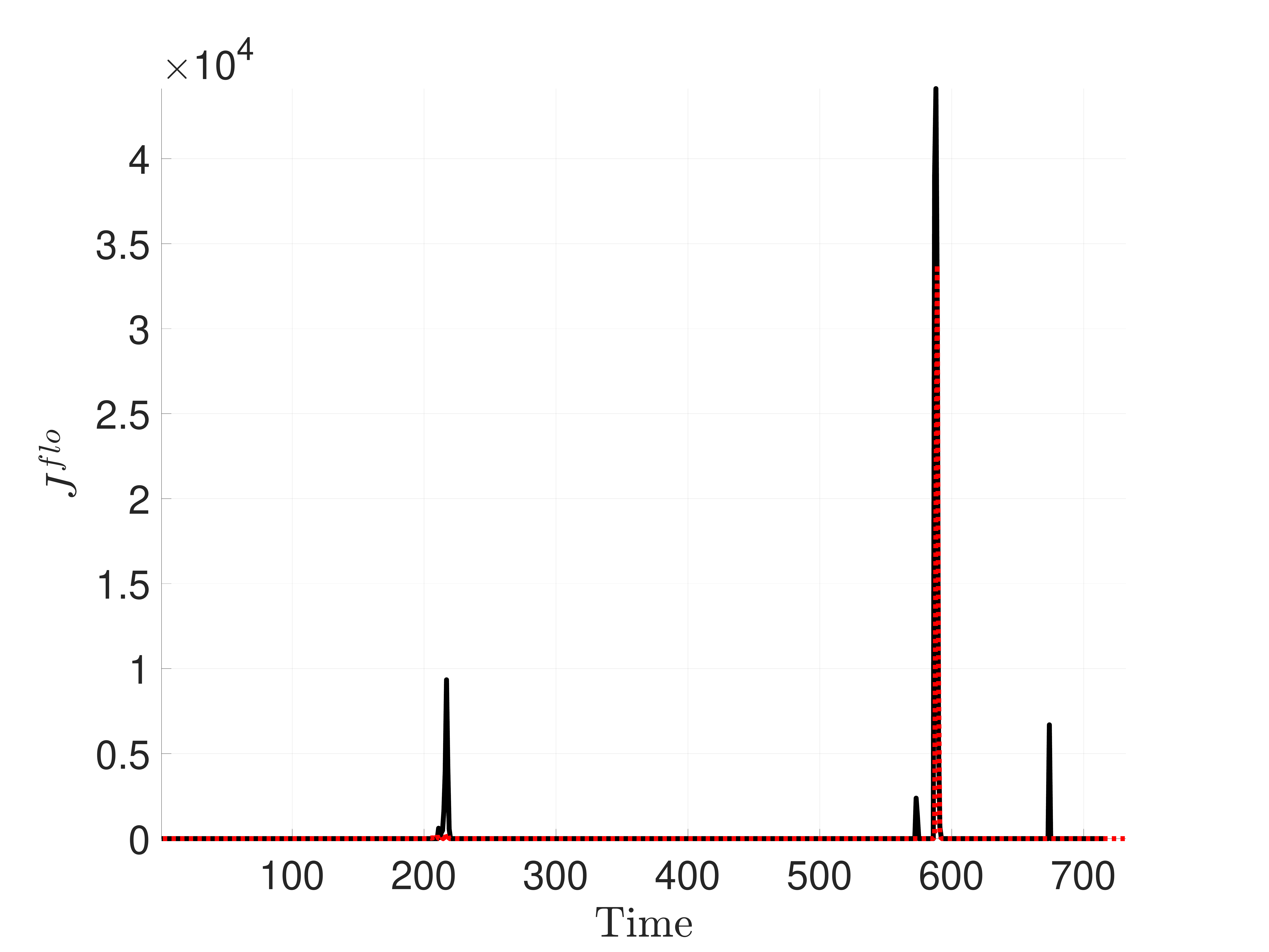}\label{Jflo_empc}}
    \caption{Performance of the controlled system in terms of hydropower production and flood damages over the period 2007-2008, compared with the DDP approach.}
    \label{fig:e_mpc_performances_2007_2008}
\end{figure*}

In order to demonstrate the ability of the proposed approach to handle different hydrologic conditions, a simulation is run for the period 2007-2008 \new{with a prediction horizon of $N_p=15$ days. The corresponding performances are visible in Figure \ref{fig:e_mpc_performances_2007_2008}, along with the optimal ones, computed by applying DDP for $\alpha=0.05$ over the 2007-2008 period: as for the training period 1962-1969, the proposed approach allows to obtain good performances in terms of hydropower production in comparison with the optimal policy, but results in higher levels of water in Hanoi. For a more realistic comparison, the proposed strategy should be compared with the performances of the policy that was obtained offline through SDP for the training period 1962-1969. The average performances over both training and validation periods are given in Table \ref{results_2007_2008} for the proposed approach (VRFT+eMPC) and the SDP strategy for $\alpha=0.05$, when trained on the period 1962-1969. %First, since the deterministic assumption does not hold anymore outside the training period, one can notice that the DDP policy can barely be reused as it leads to important floodings. 
First, the SDP policy is robust to the change of hydrologic conditions between the training and validation period, as it takes uncertainty into account by considering the disturbances as stochastic processes. On the other side, the technique proposed in this paper handles the change of hydrologic conditions by performing online optimization over a limited prediction horizon. In the end, the proposed online strategy performs the best over the validation period 2007-2008: it obtains both the highest daily average hydropower production and the lowest daily average excess level of water.}

%\new{Finally, the last line of Table \ref{results_2007_2008} corresponds to the performances that would have been obtained in 2007-2008 if applying the optimal policy $p^\star$ obtained through DDP for the previous period 1962-1969: in practice, the considered system is fed with $u^c$ under the disturbances measured in 1962-1969. Due to the evolution of the hydrologic conditions, this policy leads to much higher  the benefits of this policy are pre. This result illustrates the interest of having an online strategy that can adapt to evolving environmental conditions. }

\begin{table}[]
\centering
\begin{tabular}{c|c|c|c|c|}
\cline{2-5}
                       & \multicolumn{2}{c|}{1962-1969} & \multicolumn{2}{c|}{2007-2008} \\ \cline{2-5} 
                       & $J^H$ & $J^F$ & $J^H$ & $J^F$ \\ \hline
\multicolumn{1}{|c|}{VRFT + eMPC} & \cellcolor{green!25} 22.8049 & \cellcolor{green!25} 186.1333 & \cellcolor{green!25}23.3126 & \cellcolor{green!25} 212.9908 \\ \hline
\multicolumn{1}{|c|}{SDP} & 22.015 & 261.872 & 23.149 & 274.186 \\ \hline
%\multicolumn{1}{|c|}{DDP} & \cellcolor{green!25}25.1267 & \cellcolor{green!25}68.0440 & 22.5441 & 411.2287 \\ \hline
\end{tabular}

%\multicolumn{1}{|c|}{$u^c$ (DDP 1962-1969)} & 25.0077 & 466.6876 \\ \hline
\caption{Daily average power production and excess level of water in Hanoi for different control strategies during the validation period 2007-2008, compared with the results obtained on the training period 1962-1969 for $N_p=15$ and $\alpha=\alpha_{eMPC}=0.05$.}
\label{results_2007_2008}
\end{table}

\section{Conclusions}
\label{sec:conclusions}
In this paper, a hierarchical data-driven control design strategy is proposed for water resources management, with a focus on the Hoa Binh reservoir case study\new{, Vietnam}. First, a linear controller is designed from data (with no use of the model of the system) to approximately assign a desired behaviour to the inner-loop. The model of the inner loop is then used for the design of an outer economic model predictive control loop, aimed to handle the performance and satisfy the signals constraints.  

Compared to the resolution of the optimal control through traditional SDP tools, the proposed strategy allows to find a better compromise between hydropower production and floodings in Hanoi for a prediction horizon of 20 days. The main strength of the proposed approach is to perform online control and therefore to adapt to new environmental conditions.

Future research will include the study of the prediction of the future input flow from the different involved rivers. Moreover, many reservoirs coexist in the Red River basin, so it would be interesting to investigate their possible coordination. 

\bibliography{biblio}

\begin{thebibliography}{10}
\providecommand{\url}[1]{#1}
\csname url@samestyle\endcsname
\providecommand{\newblock}{\relax}
\providecommand{\bibinfo}[2]{#2}
\providecommand{\BIBentrySTDinterwordspacing}{\spaceskip=0pt\relax}
\providecommand{\BIBentryALTinterwordstretchfactor}{4}
\providecommand{\BIBentryALTinterwordspacing}{\spaceskip=\fontdimen2\font plus
\BIBentryALTinterwordstretchfactor\fontdimen3\font minus
  \fontdimen4\font\relax}
\providecommand{\BIBforeignlanguage}[2]{{%
\expandafter\ifx\csname l@#1\endcsname\relax
\typeout{** WARNING: IEEEtran.bst: No hyphenation pattern has been}%
\typeout{** loaded for the language `#1'. Using the pattern for}%
\typeout{** the default language instead.}%
\else
\language=\csname l@#1\endcsname
\fi
#2}}
\providecommand{\BIBdecl}{\relax}
\BIBdecl

\bibitem{mcdonald2011urban}
R.~I. McDonald, P.~Green, D.~Balk, B.~M. Fekete, C.~Revenga, M.~Todd, and
  M.~Montgomery, ``Urban growth, climate change, and freshwater availability,''
  \emph{Proceedings of the National Academy of Sciences}, vol. 108, no.~15, pp.
  6312--6317, 2011.

\bibitem{ansar2014should}
A.~Ansar, B.~Flyvbjerg, A.~Budzier, and D.~Lunn, ``Should we build more large
  dams? the actual costs of hydropower megaproject development,'' \emph{Energy
  policy}, vol.~69, pp. 43--56, 2014.

\bibitem{soncini2007integrated}
R.~Soncini-Sessa, E.~Weber, and A.~Castelletti, \emph{Integrated and
  participatory water resources management-theory}.\hskip 1em plus 0.5em minus
  0.4em\relax Elsevier, 2007.

\bibitem{castelletti2008water}
A.~Castelletti, F.~Pianosi, and R.~Soncini-Sessa, ``Water reservoir control
  under economic, social and environmental constraints,'' \emph{Automatica},
  vol.~44, no.~6, pp. 1595--1607, 2008.

\bibitem{Bellman1957}
R.~Bellman, \emph{{Dynamic programming}}.\hskip 1em plus 0.5em minus
  0.4em\relax Princeton: Princeton University Press, 1957.

\bibitem{Powell2007}
W.~Powell, \emph{{Approximate Dynamic Programming: Solving the curses of
  dimensionality}}.\hskip 1em plus 0.5em minus 0.4em\relax NJ: Wiley, 2007.

\bibitem{piga2017direct}
D.~Piga, S.~Formentin, and A.~Bemporad, ``Direct data-driven control of
  constrained systems,'' \emph{IEEE Transactions on Control Systems
  Technology}, vol.~26, no.~4, pp. 1422--1429, 2017.

\bibitem{campi2002virtual}
M.~Campi, A.~Lecchini, and S.~Savaresi, ``Virtual reference feedback tuning: a
  direct method for the design of feedback controllers,'' \emph{Automatica},
  vol.~38, no.~8, pp. 1337--1346, 2002.

\bibitem{polverini2019mixed}
M.~P. Polverini, S.~Formentin, L.~Merzagora, and P.~Rocco, ``Mixed data-driven
  and model-based robot implicit force control: A hierarchical approach,''
  \emph{IEEE Transactions on Control Systems Technology}, 2019.

\bibitem{piga2019performance}
D.~Piga, M.~Forgione, S.~Formentin, and A.~Bemporad, ``Performance-oriented
  model learning for data-driven mpc design,'' \emph{IEEE control systems
  letters}, vol.~3, no.~3, pp. 577--582, 2019.

\bibitem{faulwasser2018economic}
T.~Faulwasser, L.~Gr{\"u}ne, M.~A. M{\"u}ller \emph{et~al.}, ``Economic
  nonlinear model predictive control,'' \emph{Foundations and
  Trends{\textregistered} in Systems and Control}, vol.~5, no.~1, pp. 1--98,
  2018.

\bibitem{pianosi2012identification}
F.~Pianosi, A.~Castelletti, and M.~Lovera, ``Identification of a flow-routing
  model for the red river network,'' \emph{IFAC Proceedings Volumes}, vol.~45,
  no.~16, pp. 1037--1042, 2012.

\bibitem{Castelletti2012}
A.~Castelletti, F.~Pianosi, X.~Quach, and R.~Soncini-Sessa, ``{Assessing water
  reservoirs management and development in Northern Vietnam},'' \emph{Hydrology
  and Earth System Sciences}, vol.~16, no.~1, pp. 189--199, 2012.

\bibitem{giuliani2016curses}
M.~Giuliani, A.~Castelletti, F.~Pianosi, E.~Mason, and P.~Reed, ``Curses,
  tradeoffs, and scalable management: Advancing evolutionary multiobjective
  direct policy search to improve water reservoir operations,'' \emph{Journal
  of Water Resources Planning and Management}, vol. 142, no.~2, 2016.

\bibitem{Castelletti2012b}
A.~Castelletti, F.~Pianosi, and R.~Soncini-Sessa, ``{Stochastic and robust
  control of water resource systems: Concepts, methods and applications},'' in
  \emph{System Identification, Environmental Modelling, and Control System
  Design}.\hskip 1em plus 0.5em minus 0.4em\relax Springer, 2012, pp. 383--401.

\bibitem{Tsitsiklis1996}
J.~Tsitsiklis and B.~Van~Roy, ``{Feature-Based Methods for Large Scale Dynamic
  Programming},'' \emph{Machine Learning}, vol.~22, pp. 59--94, 1996.

\bibitem{giuliani2016large}
M.~Giuliani, D.~Anghileri, A.~Castelletti, P.~N. Vu, and R.~Soncini-Sessa,
  ``Large storage operations under climate change: expanding uncertainties and
  evolving tradeoffs,'' \emph{Environmental Research Letters}, vol.~11, no.~3,
  p. 035009, 2016.

\bibitem{formentin2019deterministic}
S.~Formentin, M.~C. Campi, A.~Car{\`e}, and S.~M. Savaresi, ``{Deterministic
  continuous-time Virtual Reference Feedback Tuning (VRFT) with application to
  PID design},'' \emph{Systems \& Control Letters}, vol. 127, pp. 25--34, 2019.

\end{thebibliography}

\end{document}